# The Duan-Kimble cavity-atom quantum memory loading scheme revisited

Michael G. Raymer,[1,*] Clark Embleton,[1] and Jeffrey H. Shapiro[2]

June 17, 2024

[1] Oregon Center for Optical, Molecular & Quantum Science, and Department of Physics,

University of Oregon, Eugene, OR 97403, USA

[2] Research Laboratory of Electronics, Massachusetts Institute of Technology,
Cambridge, Massachusetts 02139 USA

**Abstract**

We reexamine the well-known Duan-Kimble entanglement scheme, wherein the state of a single-photon qubit is entangled with a quantum memory consisting of a single-atom qubit in a strongly coupled optical cavity, providing the capability to load the photon's state into the memory. We correct a common error appearing in some subsequent papers regarding the validity of the single-photon reflectivity function that characterizes the essential phase shift at the heart of the protocol. Using the validated analytical solution, we introduce an improved scheme—the *push-pull* configuration—where the photon and cavity are tuned at the midpoint between atomic resonances and show that it can outperform the original *on-off* configuration in which the photon and cavity are tuned exactly to one of the atomic resonances. The performance metric used is the final memory-state fidelity versus the heralding probability, which determines the memory loading rate. The results should play a role in optimizing future quantum repeater schemes based on the Duan-Kimble protocol.

**Introduction**

Interfacing photonic (flying) qubits with matter (standing) qubits is a key challenge for constructing communication systems. This paper presents a nonperturbative analytical solution to a widely discussed protocol, the Duan-Kimble scheme, wherein the state of a single-photon qubit is entangled with and can be transferred into a quantum memory consisting of a single-atom (or color-center) qubit in a strongly coupled optical cavity. [[1]] The active mechanism is atomic-state-dependent cavity reflection, which performs a conditional phase shift quantum gate operation. The scheme may be used to implement a controlled phase-shift quantum gate operation between two photons in a pulse sequence [1], but here we focus on the elementary step involving a single photon packet reflecting from a cavity with an embedded atom.

Using the analytical solution, which updates previous derivations to include cavity damping and spontaneous-emission damping, enables the discovery that a *push-pull* configuration (where the photon and cavity are tuned at the midpoint between atomic resonances) can outperform the commonly discussed *on-off*, or *hot-cold*, configuration (where the photon and cavity are both tuned exactly to one of the atomic resonances). [1]

Previous solutions for this or similar setups have been either numerical [1] or perturbative and thus approximate [[2],[3],[4],[5],[6]], with the notable exceptions of Shen and Fan [[7]], Kim et. al. [[8]], and Gea-Banacloche [[9]]. which treat a closely related problem and provide insights on how to obtain analytical solutions without perturbation theory. See also related work by Mirza, van Enk, and Kimble. [[10]] The solution provided here can be used to model and optimize entanglement distribution protocols such as the recently proposed zero-added-loss multiplexing (ZALM) scheme [11] or asynchronous photonic Bell-state measurements - a key component of quantum repeaters. [2]

Of particular theoretical interest is the distinction between a 'weakly driven system' (which implies a perturbative solution would be adequate) and a 'single-quantum-excited system' (which requires a nonperturbative solution). Perhaps surprisingly, for the class of problems discussed here, the two solutions are fortuitously identical, which has led to erroneous explanations of the requirements for the widely-used solutions' validity, as perhaps first pointed out in [7]. Here we introduce an alternative and mathematically simpler solution method in which the standard input-output cavity equations, routinely derived in the Heisenberg picture (which are nonlinear and thus hard to solve), are replaced by Schrödinger-picture equations (which are linear and thus easy to solve). The present approach makes it clear why the perturbative and nonperturbative solutions lead to identical results for the class of problems discussed, consistent with the arguments given in [8] and [12].

**Model Set Up**

**Figure 1** shows three realizations of a one-sided cavity configuration. The ring cavity versions support only unidirectional modes, whereas the two-mirror version supports bidirectional or standing-wave modes. We focus on the former; the latter can be treated by a straightforward extension. Note that in **Fig.1** (a) and (b) we do not consider reflection in the backward direction opposite to the input $A(t)$, whereas such is considered in some other treatments such as [7]. We term the field $B(t)$ as 'reflection' in all three configurations shown; this terminology and the associated 'reflection' coefficients vary among the papers cited but are consistent with our earlier work [13] upon which this paper builds.

Cavity input-output theory merges the continuum of modes external to a cavity with the discrete set of 'quasi-modes' inside the cavity and has been treated using a wide range of approaches. [14, 15 7, 10] Typically such treatments allow for only a single cavity mode, assuming a large free-spectral range (FSR) separating the mode frequencies in a very high-finesse cavity. Raymer and McKinstrie generalized the treatment to allow for many longitudinal modes, which are naturally excited when an ultrashort light pulse is incident on a cavity. [13] That approach, which also allows for low cavity finesse, has been used, for example, for modeling Raman-based atomic quantum memories. [16] The input-output formalism for cavities can be viewed as a theory of mode transformations rather than state transformations, and thus applies equally well to classical or quantum problems when treated in the Heisenberg picture.

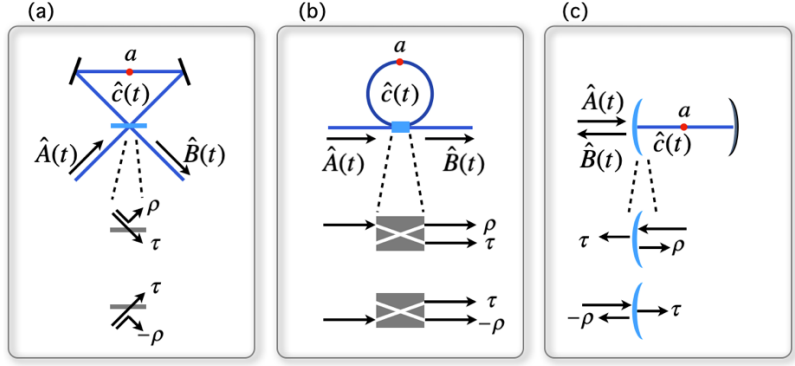

Fig.1. (a), (b), and (c) show three mathematically equivalent single-sided cavity configurations. $A(t)$ input, $B(t)$ output, and $c(t)$ intra-cavity are optical fields written with carets to indicate quantum field operators. The amplitude reflection $\rho$ and transmission $\tau$ coefficients are shown with the chosen sign convention. The single atom or color center is indicated by $a$. Adapted from [13].

**Figure 2** shows the application of the Duan-Kimble scheme we emphasize in this paper. The model for the atom is shown in **Fig. 3** and assumes a ground-state separation of frequency $\Delta$, giving two independent dipole-allowed transitions, as appropriate for, e.g., an atom with suitable selection rules or a silicon-vacancy center in diamond (with an appropriate magnetic field applied). [2,17] Here we treat the simplest case that the light entering the cavity is in a single, known polarization state (typically horizontal or vertical) and that the two transitions are uncoupled. (In contrast, for exciton transitions in a semiconductor quantum dot, angular-momentum selection rules lead to coupled transitions for light in arbitrary polarization states, leading to effects such as Faraday rotation. [6,8])

A single-photon wave packet is initially in the polarization state $\alpha|H\rangle + \beta|V\rangle$ (or equivalently the time-bin encoded state $\alpha|early\rangle + \beta|late\rangle$). The goal is to transfer this state into the ground states $|g_1\rangle, |g_2\rangle$ of the atom, which is prepared initially in the superposition $2^{-1/2}(|g_1\rangle + |g_2\rangle)$. The photon packet is split by a polarizing beam splitter (PBS), with the $H$ component reflecting from the cavity acquiring an atomic-state-dependent phase shift, then is rotated by a half-wave plate to have $V$ polarization. The $V$ component reflects from a mirror with path length adjusted to provide zero phase shift, whereupon the two packets are interfered at a 50/50 beam splitter, and the atom is transformed by a Hadamard operation. The system is designed such that ideally the reflected packet is not distorted temporally but acquires a conditional phase shift whose difference equals $\pi$ radians depending on which atomic ground state the photon interacts with. After the interaction, the joint photon-atom state is entangled. Then, depending on which detector registers an event, the atomic state is projected into the targeted state $\alpha|g_1\rangle \pm \beta|g_2\rangle$, the sign of $\beta$ being known and easily corrected by a subsequent unitary operation if desired.

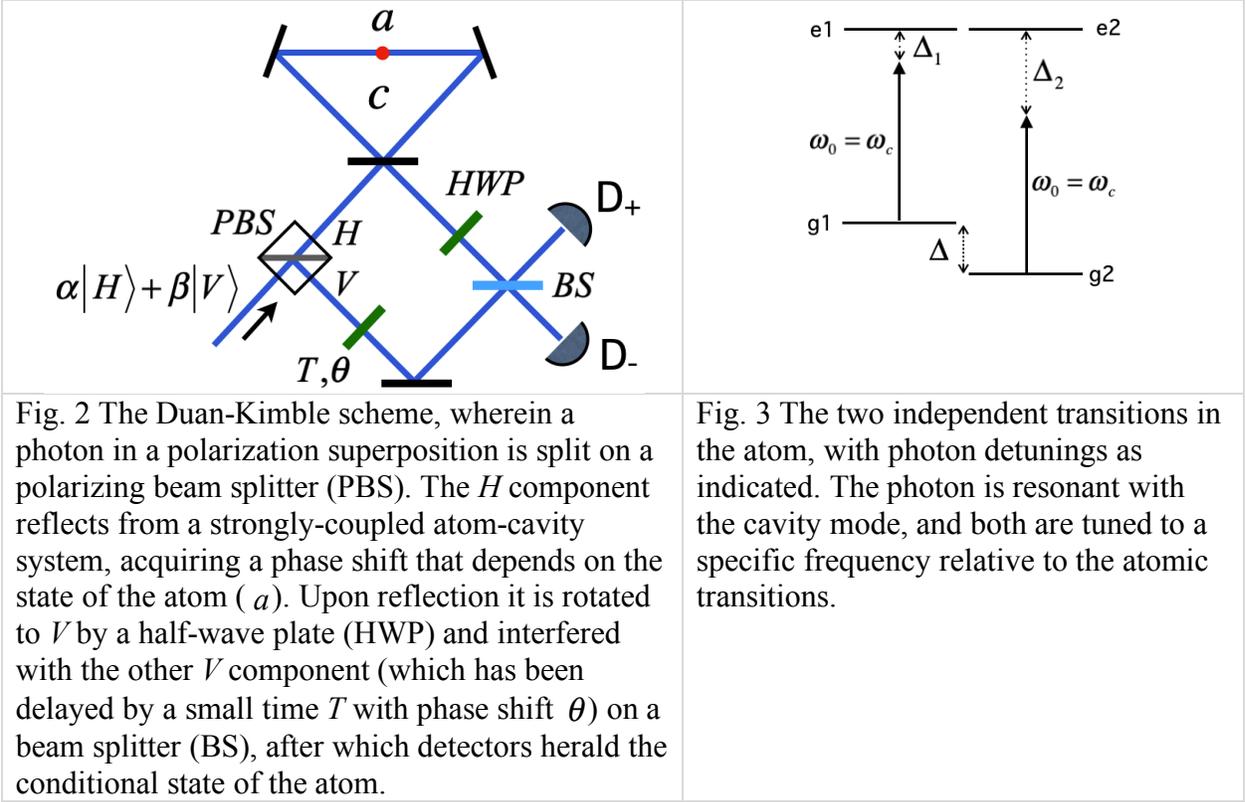

Fig. 2 The Duan-Kimble scheme, wherein a photon in a polarization superposition is split on a polarizing beam splitter (PBS). The $H$ component reflects from a strongly-coupled atom-cavity system, acquiring a phase shift that depends on the state of the atom ($a$). Upon reflection it is rotated to $V$ by a half-wave plate (HWP) and interfered with the other $V$ component (which has been delayed by a small time $T$ with phase shift $\theta$) on a beam splitter (BS), after which detectors herald the conditional state of the atom.

Fig. 3 The two independent transitions in the atom, with photon detunings as indicated. The photon is resonant with the cavity mode, and both are tuned to a specific frequency relative to the atomic transitions.

**Cavity-Atom Input-Output Theory**

To model cavity transmission and/or scattering losses, we add cavity coupling to a vacuum input field $J_{in}(t)$ and output field $J_{out}(t)$. Relevant cavity coupling rates (one-half of energy decay rates) are denoted $\kappa, \kappa_Q, \kappa_J$. To include loss by spontaneous emission loss by the atom into modes other than the $c$ cavity, we couple the atom's dipole to a fictitious second cavity, with field $q(t)$, which in turn is coupled to a vacuum input field $Q_{in}(t)$ and output field $Q_{out}(t)$, and we eliminate the $Q(t)$ cavity in later steps.

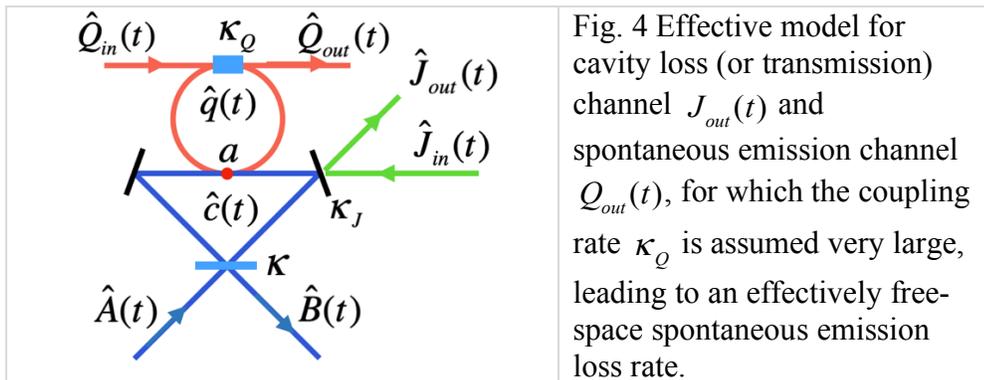

Fig. 4 Effective model for cavity loss (or transmission) channel $J_{out}(t)$ and spontaneous emission channel $Q_{out}(t)$, for which the coupling rate $\kappa_Q$ is assumed very large, leading to an effectively free-space spontaneous emission loss rate.

A unidirectional propagating photon-annihilation operator (positive-frequency part) is defined in **Appendix A** and denoted $\hat{\Phi}^{(+)}(t,z)$. When evaluated at the input port ($z = 0_-$) and at the output port ($z = 0_+$) of the cavity, it defines slowly-varying input and output photon-field operators

$$\hat{A}(t) = \int_{-\omega_0}^{\infty} \frac{d\omega}{2\pi} \hat{a}_{in}(\omega) e^{-i\omega t}$$
$$\hat{B}(t) = \int_{-\omega_0}^{\infty} \frac{d\omega}{2\pi} \hat{a}_{out}(\omega) e^{-i\omega t}, \quad (1)$$

through the relations $\hat{\Phi}^{(+)}(t,z=0_-) = \hat{A}(t)e^{-i\omega_0 t}$ and $\hat{\Phi}^{(+)}(t,z=0_+) = \hat{B}(t)e^{-i\omega_0 t}$, with $\omega$ being the detuning from the carrier frequency $\omega_0$. The fields are assumed to be narrow-band relative to the carrier frequency and are defined such that, for example, $\hat{A}^\dagger(t)\hat{A}(t)$ is a photon flux operator localized at the point of entry to the cavity. (We suppress polarization labels here, as we assume the polarization is not affected by the cavity interaction.) Analogous relations hold for $\hat{J}_{in}(t), \hat{J}_{out}(t)$ and $\hat{Q}_{in}(t), \hat{Q}_{out}(t)$.

For the problem being considered, the state of the optical field at the input to the cavity is a single-photon state with a well-defined polarization and 'temporal mode,' as reviewed in [18], defined in the frequency domain as

$$|\varphi\rangle_{in} = \int \frac{d\omega}{2\pi} \tilde{A}(\omega) \hat{a}_{in}^{\dagger}(\omega) |vac\rangle, \quad (2)$$

where the integral is understood to extend over the support of the square-normalized spectral amplitude function $\tilde{A}(\omega)$ that defines the form of the wave packet. The cavity output state is, similarly,

$$|\varphi'\rangle_{out} = \int \frac{d\omega}{2\pi} \tilde{B}(\omega) \hat{a}_{out}^{\dagger}(\omega) |vac\rangle. \quad (3)$$

Because energy can be lost from the subsystem during the cavity interaction, this state is not necessarily normalized. The initial task at hand is to learn the mapping from input spectral amplitude $\tilde{A}(\omega)$ to output spectral amplitude $\tilde{B}(\omega)$.

In the case of a single cavity mode being excited (input spectrum being much narrower than the free spectral range of the cavity), the quasi-mode annihilation operator for the cavity satisfies the equal-time commutator $[\hat{c}(t), \hat{c}^\dagger(t)] = 1$ and likewise for $\hat{q}(t)$.

Under the assumption of high cavity finesse ($\rho \approx 1$), all the variants of input-output theory lead to the same input-output operator equations [13,14,15],

$$\hat{B}(t) = -\hat{A}(t) + \sqrt{2\kappa}\,\hat{c}(t)$$
$$\hat{J}_{out}(t) = -\hat{J}_{vac}(t) + \sqrt{2\kappa_J}\,\hat{c}(t) \quad (4)$$
$$\hat{Q}_{out}(t) = -\hat{Q}_{vac}(t) + \sqrt{2\kappa_Q}\,\hat{q}(t),$$

with the phase convention for reflection conforming to that in [13].

Consider a single two-level atomic transition being present (with frequency $\omega_{eg}$), coupled to cavities $C$ and $Q$ with single-photon Rabi frequencies $g$ and $g_Q$ respectively. The Heisenberg-picture equations of motion, including operator Bloch-Langevin equations for the atom in the rotating-wave approximation (valid for $g, g_Q \ll \omega_0$), are expressed in terms of field operators and the atomic transition operator, which at the initial time is $\hat{\sigma}_{ge} = |g\rangle\langle e|$, and the population projection operator, initially $\hat{\sigma}_{ee} = |e\rangle\langle e|$. All variables are defined as slowly varying variables by factoring out the rapidly oscillating term $exp(-i\omega_0 t)$, leading to the standard Heisenberg-picture equations, (e.g., [19])

$$\partial_t \hat{c}(t) = -(i\Delta_c + \kappa + \kappa_J)\hat{c}(t) + g\hat{\sigma}_{ge}(t) + \sqrt{2\kappa}\,\hat{A}_{in}(t) + \sqrt{2\kappa_J}\,\hat{J}_{in}(t)$$
$$\partial_t \hat{q}(t) = -\kappa_Q \hat{q}(t) + g_Q \hat{\sigma}_{ge}(t) + \sqrt{2\kappa_Q}\,\hat{Q}_{in}(t)$$
$$\partial_t \hat{\sigma}_{ge}(t) = -i\Delta_e \hat{\sigma}_{ge}(t) + g\hat{c}(t)(\hat{\sigma}_{ee}(t) - \hat{\sigma}_{gg}(t)) + g_Q \hat{q}(t)(\hat{\sigma}_{ee}(t) - \hat{\sigma}_{gg}(t)) \quad (5)$$
$$\partial_t \hat{\sigma}_{ee}(t) = -g\hat{\sigma}_{eg}(t)\hat{c}(t) + g\hat{c}^\dagger(t)\hat{\sigma}_{ge}(t) - g_Q \hat{\sigma}_{eg}(t)\hat{q}(t) + g_Q \hat{q}^\dagger(t)\hat{\sigma}_{ge}(t).$$

The cavity resonance frequencies are $\omega_c, \omega_q$ and detunings are defined as $\Delta_c = \omega_c - \omega_0$ and $\Delta_e = \omega_{eg} - \omega_0$, and it is assumed that the spontaneous-emission cavity's frequency is resonant with the atomic transition frequency, $\omega_q = \omega_{eg}$. Note that all damping coefficients correspond to amplitude damping, whereas some other papers define them as energy damping rates, which are twice the values of ours. We have neglected any pure dephasing interactions (such as dephasing atomic or phonon interactions [20]), so there is no need to add additional Langevin noise operators, i.e., the system is unitary at this level. We will treat the system as if the atom is in one or the other of its ground states, so it is sufficient to treat the two cases separately.

The operator equations contain nonlinear contributions, making their full solution problematic in the general case. Two approaches have been followed in the literature to arrive at linear equations, which are easy to solve. The original proposal of Duan and Kimble used numerical methods to solve the corresponding Schrödinger-picture equations, whereas the present method allows solving the same problem analytically. Other papers solved operator equations similar to Eq.(5) by making the (unwarranted) assumption that the ensemble-averaged excited-state population $\langle \hat{\sigma}_{ee}(t) \rangle$ remains much less than one, which is not the case in a strongly-coupled atom-cavity interaction, as we verify below. The error was pointed out and remedied in [7] and [8], for slightly different problems and using different methods than in the present paper. (As mentioned above, this assumption leads fortuitously to the same result as the exact method presented here, as proved below.) An additional alternative method to solve such problems is quantum trajectory theory. [21]

In **Appendix A** we review how to transform Eq.(5) to the Schrödinger picture in cases where only one quantum excitation is present in the entire system. Closely related derivations have been carried out previously, as described in [9], but none treat the full problem with all loss channels as needed here. To sketch our method for a simple case omitting the $q$ cavity and the $J$ field channels, we write the pure state for the atom, the cavity field $c$, and combined exterior field ($A$ and $B$), respectively, as

$$|\psi(t)\rangle = \psi_e(t)|e\rangle_a|0\rangle_c|vac\rangle_{AB} + \psi_c(t)|g\rangle_a|1\rangle_c|vac\rangle_{AB} + \int_0^\infty \frac{dk}{2\pi} \psi_k(t)|g\rangle_a|0\rangle_c|1_k\rangle_{AB}, \qquad (6)$$

where $\psi_e(t)$ and $\psi_c(t)$ are the amplitudes for the excitation to be in the atom or in the cavity $C$. The basis states are $|j\rangle_{atom}|n\rangle_c|vac\rangle_{AB}$ or $|j\rangle_{atom}|n\rangle_c|1_k\rangle_{AB}$ with $j = e, g$ for the atomic state, $n = 0, 1$ for the cavity photon number, $|vac\rangle_{AB}$ indicating no photons in the exterior field, and $|1_k\rangle_{AB}$ indicating one photon in the exterior field with frequency $(\omega_0 + \omega)$ and wave number $k(\omega) = (\omega_0 + \omega)/c$, $c$ being the speed of light. We define $t_p$ to be an arbitrary time far in the past before the single-photon packet arrives at the cavity, at which time the initial conditions are $\psi_e(t_p) = \psi_c(t_p) = 0$. The Schrödinger equation is derived using a Hamiltonian given in **Appendix A**, and standard approximations akin to the Wigner-Weiskopf approximation are used to derive the set of equations (now including decay into the $J_{out}$ channel as well as the $Q$ cavity and its decay into the $Q_{out}$ channel)

$$\partial_t \psi_e(t) = -i\Delta_e \psi_e(t) - g\psi_c(t) - g_Q \psi_q(t)$$
$$\partial_t \psi_c(t) = -(i\Delta_c + \kappa + \kappa_J)\psi_c(t) + g\psi_e + \sqrt{2\kappa}\, A(t) \qquad (7)$$
$$\partial_t \psi_q(t) = -\kappa_Q \psi_q(t) + g_Q \psi_e(t).$$

where $\psi_q(t)$ is the amplitude for the excitation to be in the $q$ cavity and parameters are the same as before. The rapidly oscillating factor $\exp(-i\omega_0 t)$ has been factored out of all the amplitudes in Eq.(7), so they are slowly varying. $A(t)$ is the quantum amplitude of the single-photon state occupying the temporal mode entering the cavity, expressed in the time domain as

$$A(t) = \int_{-\omega_0}^{\infty} \frac{d\omega}{2\pi} \tilde{A}(\omega) e^{-i\omega t}, \qquad (8)$$

where $\tilde{A}(\omega) = c^{-1/2} \psi_{k(\omega)}(t_p) e^{i\omega t_p}$ is the square-normalized spectral amplitude before the interaction. For Eq.(7) we assumed there is no light entering the $J_{in}$ and $Q_{in}$ channels. The input-output relation derived (**Appendix A**) in the Schrödinger picture yields for the time-domain output amplitude

$$B(t) = -A(t) + \sqrt{2\kappa}\, \psi_c(t) , \qquad (9)$$

corresponding to the frequency-domain output state amplitude $\tilde{B}(\omega) = -c^{-1/2} \psi_{k(\omega)}(t_f) e^{i\omega t_f}$, where the minus sign assumes a particular phase convention for reflection and $t_f$ is an arbitrary far-future time well after the photon packet exits the cavity ($c$ is again the speed of light). Fourier transforming we have

$$\tilde{B}(\omega) = -\tilde{A}(\omega) + \sqrt{2\kappa}\, \tilde{\psi}_c(\omega) . \qquad (10)$$

Likewise, we can derive for the cavity loss channel,

$$\tilde{J}_{out}(\omega) = \sqrt{2\kappa_J}\, \tilde{\psi}_c(\omega) . \qquad (11)$$

Equations (8) and (9) are seen to be consistent with Eqs.(2), (3), and (4) when noting that Eq.(2) is equivalent to

$$\left|\varphi\right\rangle_{in} = \int \frac{d\omega}{2\pi} \tilde{A}(\omega) \left|1_{k(\omega)}\right\rangle , \qquad (12)$$

matching the last term in Eq.(6). The input-output formalism is a scattering theory that relates the scattered spectral amplitude $\psi_\omega(t_f)$ to the incident spectral amplitude $\psi_\omega(t_p)$.

We note that the Schrödinger-picture equations are linear, as desired. The nonlinearity is removed because in a pure-state formalism, the population of the atomic excited state is fully accounted for by the mod-square of the amplitude $\psi_e(t)$ (but note that the formalism as it stands does not allow for pure dephasing, such as by elastic atomic collisions or phonon scattering.

To find the relation between input and output amplitudes, we eliminate the spontaneous emission cavity ($Q$) by taking $\kappa_Q$ to be much larger than any other relevant rate, so the cavity acts like a unidirectional free-space loss channel. Then the limit of large $\kappa_Q$ gives $\psi_q(t) \approx (g_Q / \kappa_Q) \psi_e(t)$ and thus,

$$\begin{aligned}\partial_t \psi_c(t) &= -(i\Delta_c + \kappa + \kappa_J)\psi_c(t) + g\psi_e + \sqrt{2\kappa}\, A(t) \\ \partial_t \psi_e(t) &= -(i\Delta_e + \gamma)\psi_e(t) - g\psi_c(t) .\end{aligned} \qquad (13)$$

where the spontaneous damping rate is $\gamma = g_Q^2 / \kappa_Q$, which is essentially Fermi's Golden Rule. (This 'bad-cavity' limit is also discussed in [9].)

A key point is that Eq.(7) is identical to what one obtains by considering the expectation values of each term in Eq.(5) and (incorrectly) assuming that the population $\langle \hat{\sigma}_{ee} \rangle$ remains much less than one, as done in several publications. The present derivation shows why the (incorrect) perturbation theory fortuitously gives the correct result.

To cement our argument, note that when Eq.(13) is solved numerically as a function of time, it shows that the atomic excited-state population does *not* remain much less than one, invalidating the perturbative solution method that has been assumed valid in some previous papers, as mentioned earlier. For example, with a normalized single-photon Gaussian input pulse of duration 1 (in arbitrary units), and parameters $\kappa = 1, g = 1, \gamma = 0.01$ and $\kappa_J = 0$, we find that the excited-state population $|\psi_e(t)|^2$ exceeds 0.8 at its peak.

Here we are more interested in the solution for the reflected field after it completes its interaction with the cavity. The solution is best found in the frequency domain, keeping in mind that the functions vanish at very early and very late times, so the Fourier transform of all functions, $\tilde{f}(\omega) = \int f(t) exp(i\omega t) dt$, yields the needed information,

$$-i\omega \tilde{\psi}_e(\omega) = -(i\Delta_e + \gamma)\tilde{\psi}_e(\omega) - g\tilde{\psi}_c(\omega)$$
$$-i\omega \tilde{\psi}_c(\omega) = -(i\Delta_c + \kappa + \kappa_J)\tilde{\psi}_c(\omega) + g\tilde{\psi}_e(\omega) + \sqrt{2\kappa}\,\tilde{A}(\omega). \quad (14)$$

Solving these linear equations for $\tilde{\psi}_c(\omega)$ and using Eq.(10), the solution for the spectral amplitude of the reflected photon state is

$$\tilde{B}(\omega) = r_j(\omega)\tilde{A}(\omega), \quad (15)$$

where, depending on which state the atom is in ( $j = 1$ or $2$ ), the complex reflection coefficient is

$$r_j(\omega) = \frac{(\gamma + i\Delta_j - i\omega)(\kappa - \kappa_J - i\Delta_c + i\omega) - g^2}{(\gamma + i\Delta_j - i\omega)(\kappa + \kappa_J + i\Delta_c - i\omega) + g^2}$$
$$= \frac{(1 + i\Delta_j/\gamma - i\omega/\gamma)(1 - \kappa_J/\kappa - i\Delta_c/\kappa + i\omega/\kappa) - C}{(1 + i\Delta_j/\gamma - i\omega/\gamma)(1 + \kappa_J/\kappa - i\Delta_c/\kappa - i\omega/\kappa) + C}. \quad (16)$$

The 'single-atom lossless cooperativity' is defined here as $C = g^2/\kappa\gamma$ (the cavity loss rate $\kappa_J$ is not accounted for in this definition). Note that various other studies sometimes include $\kappa_J$ and either multiply or divide by 2 in alternate definitions of $C$. Recall that $\omega$ is the detuning from the carrier frequency $\omega_0$. It's worth noting that for the *push-pull* configuration, where $\Delta_1 = -\Delta_2$, if the photon and cavity have the same center frequencies ($\Delta_c = 0$), then the reflection coefficient satisfies the symmetry $r_2(-\omega) = r_1^*(\omega)$. The *on-off* configuration does not have this symmetry. Although we have treated only the simplest scenario, the method used is easily generalized to more complex scenarios: bidirectional cavity modes, coupled atomic transitions, two or more excitations in the system, etc.

If there are no losses ($\gamma = \kappa_J = 0$), then the magnitude of $r_j(\omega)$ equals 1 and all light entering the cavity is passed into the $B$ output channel. As mentioned earlier, the result Eq.(16) is identical to that given in several papers based on the perturbative weak-excitation approximation, although such agreement is fortuitous (and fortunate for those who have used it).

A useful limit is when the photon is very narrow-band and tuned to the cavity resonance ($\Delta_c = 0$). Then the amplitude reflection coefficient is well approximated by its value at $\omega = 0$,

$$r_j(0) = \frac{(1+i\Delta_j/\gamma)(1-\kappa_J/\kappa) - C}{(1+i\Delta_j/\gamma)(1+\kappa_J/\kappa) + C} \quad . \tag{17}$$

Having a validated solution in hand, we can analyze different memory-loading scenarios and evaluate their state fidelity.

**Conditional Reflectivities and Phase Shifts**

Here we compare two schemes for optimizing the differential phase shift (to be nearly $\pi$ radians) while using the lowest cooperativity possible and with the least amount of wave-packet reshaping, thus achieving high fidelity of atomic memory loading. The initially proposed, and commonly cited, scheme we call the *on-off* scheme, wherein one of the atomic resonances is on-resonance with the cavity (and the photon), and the other is far-off resonance. That is, $\Delta_1 = 0, \Delta_2 = \Delta$, where $\Delta$ is the ground-state separation (assuming the upper states are degenerate, as in Fig. 3). The far-off-resonance case ($\Delta_2 = \Delta$) is often idealized as an empty, or 'cold,' cavity having the standard bare-cavity reflective phase shift. The on-resonance case ($\Delta_1 = 0$) leads to a dressed-atom Rabi splitting of the coupled cavity-and-atom system, creating resonances at $\pm g$ and ideally a reflected wave packet that is $\pi$ out of phase with the bare-cavity case.

We introduce the *push-pull* configuration, where the photon and cavity are tuned at the midpoint between atomic resonances ($\Delta_2 = -\Delta_1 = \Delta/2$), resulting in dressed atom-cavity resonances shifted by the ac Stark effect with frequencies $\pm\sqrt{(\Delta/2)^2 + g^2}$, For an optimized value of the cooperativity, the two atomic-state cases contribute equal and opposite phase shifts of $\pm\pi/2$ at zero detuning. (The ideal reflectivities are then $\pm i$.) Compared to the *on-off* scheme, this scheme offers higher memory-loading fidelity while maintaining high loading probability, as we will show.

For either case, the phase-shift difference upon reflection is frequency-dependent according to

$$\delta_{phase}(\omega) = \theta_1(\omega) - \theta_2(\omega) , \tag{18}$$

where the atomic-state-dependent phases are

$$\theta_j(\omega) = Arg\left[r_j(\omega)\right] \ (j=1,2) . \tag{19}$$

For the *push-pull* scheme, assuming $\Delta_c = 0$, we can solve for the value of $C$ that produces exactly the desired phase-shift difference at center frequency $\delta_{phase}(0) = \pi$, and is thus most relevant for a photon with a very small bandwidth. On the condition that $\kappa_J \leq \kappa$, the solution is

$$C_\pi = \sqrt{1 + \left(1 - \frac{\kappa_J^2}{\kappa^2}\right)\frac{\Delta_1^2}{\gamma^2}} - \frac{\kappa_J}{\kappa} . \tag{20}$$

For $C = C_\pi$, the complex amplitude reflectivity at center frequency, Eq.(17), becomes pure imaginary,

$$r_j(0) = i\Delta_j \frac{\sqrt{\gamma^2\kappa^2 + \Delta_j^2\left(\kappa^2 - \kappa_J^2\right)} - \gamma\kappa}{\Delta_j^2\left(\kappa + \kappa_J\right)} \quad (j=1,2). \tag{21}$$

Thus, because $\Delta_2 = -\Delta_1$, the reflected field (at center frequency) acquires a $+i$ or $-i$ factor (that is, a relative $\pi$ phase difference) depending on which atomic ground state is occupied, as desired. Equation (21) also shows that it's important to maintain low cavity losses ($\kappa_J \ll \kappa$) because for $C = C_\pi$ the reflectivity (at center frequency) approaches zero as the value $\kappa_J$ approaches that of $\kappa$.

In contrast, for the *on-off* scheme, we find there is no condition for which the value of $C$ produces a phase difference value exactly $\pi$ because the *off* condition (that the atom is not coupled at all to the cavity field) is never perfectly satisfied. (In one experiment the atom was physically removed from the cavity, allowing the *off* condition to be satisfied [4].) We also note that, unlike in the *push-pull* scheme, for the *on-off* scheme the cavity reflectivities at zero detuning (Eq.(17)) for the two atomic states are not equal in magnitude, which can compromise the memory-loading fidelity.

Here we examine conditions for optimizing the phase difference in the *on-off* scheme. For this, we assume the atom-cavity detuning to be large ($\Delta_2 \gg \gamma$) to minimize the atom's effect on the field. See **Appendix B** for details. First consider an ideal cavity having no losses ($\kappa_J = 0$). In this case, the energy reflectivity equals nearly 1 at all frequencies when the spontaneous emission rate $2\gamma$ is small compared to $\kappa$. To reach a phase difference between *on* and *off* cases as close as possible to $\pi$, one needs to satisfy the conditions, assuming $\Delta_1 = 0$,

$$1 \ll \frac{g^2}{\gamma\kappa} \ll \frac{\Delta_2}{\gamma} , \tag{22}$$

or in terms of the cooperativity, $1 \ll C \ll \Delta_2/\gamma$. Under these conditions, the phase-shift error $\delta_{error}$ (departure of $\delta_{phase}(0)$ from $\pi$) for the *off* condition is estimated and lower bounded by the expressions

$$\delta_{error} \approx \frac{g^2}{\gamma\kappa}\frac{2\gamma}{\Delta_2} = C\frac{2\gamma}{\Delta_2} \gg \frac{2\gamma}{\Delta_2} , \tag{23}$$

which can be very small but never zero. If cavity loss is small ($\kappa_J \ll \kappa$) but not zero, the estimate is slightly altered to

$$\delta_{error} \approx C\frac{2\gamma}{\Delta_2(1-\kappa_J^2/\kappa^2)} \gg \frac{2\gamma}{\Delta_2} . \tag{24}$$

It is desirable that the energy reflectivity of the photon packet be close to 1 to maximize the probability of a successful heralding event. Thus, we define the energy (probability) reflectivities $R_j$ as

$$R_j = \int_{-\infty}^{\infty} \frac{d\omega}{2\pi} |\tilde{A}(\omega)|^2 |r_j(\omega)|^2 , \tag{25}$$

and, for given $\alpha$ and $\beta$, the probability of observing a heralding event at either detector is

$$P_{herald} = |\beta|^2 + |\alpha|^2 \left(\frac{R_1 + R_2}{2}\right), \tag{26}$$

which, in the 'worst case' ($\beta = 0$), equals $(R_1 + R_2)/2$.

In **Fig. 5** we illustrate the operating principles of the two schemes. Parts (a) and (b) correspond to the *on-off* scheme with $\Delta = 100\kappa$, $\sigma = \kappa$, $\gamma = \kappa$, $\kappa_J = 0, C = 10$, and the horizontal axis being scaled as $\delta\omega = \omega/\kappa$. (a) is with the atom in state 1, which is far-off resonance so the reflectivity (shown as solid and dashed for the absorptive and dispersive components $\text{Re}\,r_j(\omega)$ and $\text{Im}\,r_j(\omega)$

) have structure around zero frequency. (b) is with the atom in state 2, which is on-resonance so the reflectivity is ac-Stark split into a doublet with peaks at $\pm g/\kappa = \pm 3.26$. For each case an orange dot marks the value of the phase shift at zero frequency (where the incident photon spectrum is assumed to be concentrated), with value zero in case (a) and approximately $\pi$ in case (b), giving the desired difference close to $\pi$.

Parts (c) and (d) correspond to the *push-pull* scheme with $\Delta = 100\kappa$, $\sigma = \kappa$, $\gamma = \kappa$, $\kappa_J = 0, C = C_\pi = 50.01$. The cavity and photon frequencies lie midway between the atomic resonances. (c) is with the atom in state 1, and the phase shift is $-\pi/2$, while in (d) the atom in state 2, and the phase shift is $+\pi/2$, giving the desired difference of $\pi$.

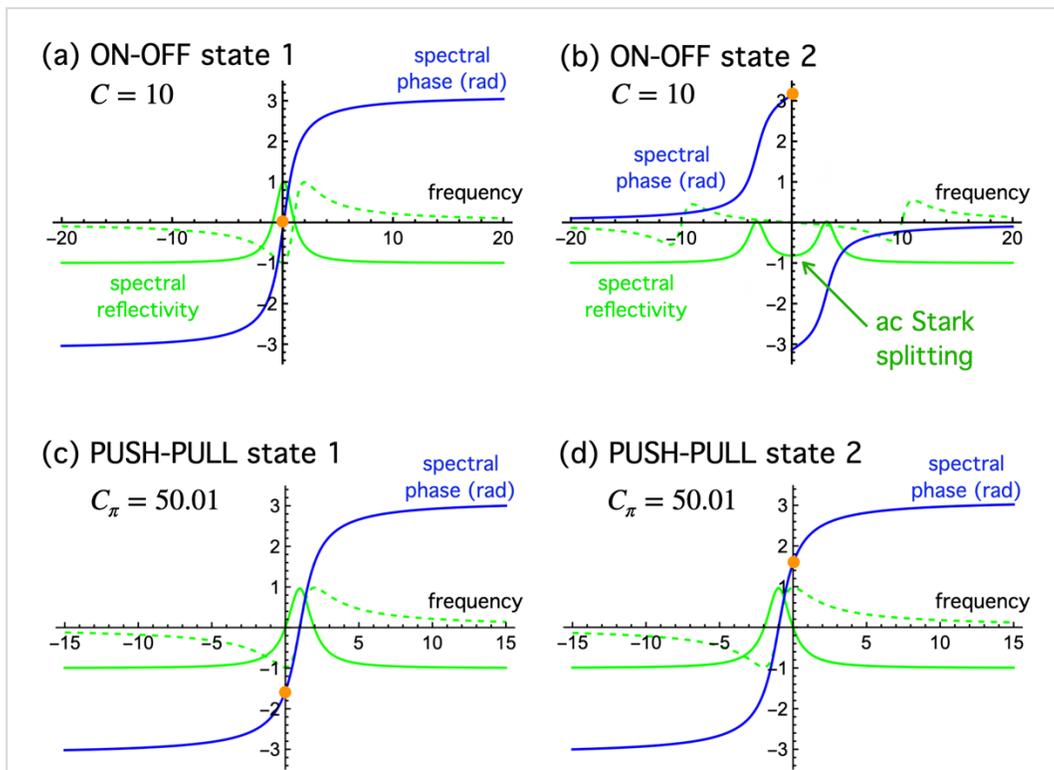

Fig.5. Operating principles of (a, b) *on-off* and (c, d) *push-pull* cavity reflection schemes. See text for description. The absorptive and dispersive components of the atomic response are shown in green (solid and dashed, respectively). The discontinuities in the spectral phase-shift curve in (b) (blue) is merely from the plotting routine confining the values to be within $\pm\pi$.

**Atomic State Fidelity**

We define the 'memory fidelity' as the overlap of the resulting normalized atomic state $\hat{\rho}_s$ with the ideal targeted state $\alpha|g_1\rangle + s\beta|g_2\rangle$, depending on which detector registers a photon (determining the sign $s = \pm 1$). The fidelity is thus

$$F_s = \left\{ \alpha^* \langle g_1| + s\beta^* \langle g_2| \right\} \hat{\rho}_s \left\{ \alpha|g_1\rangle + s\beta|g_2\rangle \right\}. \tag{27}$$

As shown in **Appendix C**, we find.

$$F_s = \frac{1}{8K_s} \int \frac{d\omega}{2\pi} |\tilde{A}(\omega)|^2 \left| \left( |\alpha|^2 + s\beta^*\alpha \right) r_1(\omega) - \left( |\alpha|^2 - s\beta^*\alpha \right) r_2(\omega) + 2|\beta|^2 e^{i\theta} e^{i\omega T} \right|^2, \tag{28}$$

where $\theta$ and $T$ are adjustable phase and delay parameters for fidelity optimization, as in **Fig.2**, and the normalization constant is

$$K_s = \frac{1}{8} \int \frac{d\omega}{2\pi} |\tilde{A}(\omega)|^2 \left\{ |\alpha(r_1(\omega) - r_2(\omega))|^2 + |\alpha[r_1(\omega) + r_2(\omega)] + 2s\beta e^{i\theta} e^{i\omega T}|^2 \right\}. \tag{29}$$

For ideal operation we would have $r_1(\omega) = -r_2(\omega) = e^{i\theta} e^{i\omega T}$ across the entire spectral range of the incoming photon, leading to $F_s = 1$, as desired.

**Comparisons of *push-pull* and *on-off* performance**

Here we compare the performance of the *push-pull* scheme (where the photon and cavity are tuned at the midpoint between atomic resonances) and *on-off* scheme (where the photon and cavity are tuned to one of the atomic resonances) for various parameter values, including values close to those used in the original Duan-Kimble paper that proposed the *on-off* scheme. The square-normalized incident wave packet is modeled as $\tilde{A}(\omega) = (8\pi/\sigma^2)^{1/4} \exp(-\omega^2/\sigma^2)$, with linewidth parameter $\sigma \ll \omega_0$. We assume the photon is tuned to the cavity resonance ($\Delta_c = 0$).

The memory fidelity depends on the magnitudes and relative phase of state parameters $\alpha, \beta$ of the incoming photon, whereas we wish to have a measure of fidelity that is independent of the incoming state. A common scenario is that the unknown incoming photon has equal probability to be in a polarization state anywhere on its Poincaré (Bloch) sphere. Thus, representing the qubit state $\alpha|g_1\rangle + \beta|g_2\rangle$ in spherical coordinates with

$$(\alpha, \beta) = \left(\cos(\chi/2), e^{i\phi}\sin(\chi/2)\right), \tag{30}$$

we define an amplitude-and-phase-averaged fidelity as

$$F_{ave,s} = \frac{1}{4\pi} \int_{-\pi}^{\pi} d\phi \int_{0}^{\pi} \sin\chi \, d\chi \, F_s, \tag{31}$$

and, likewise, the amplitude-and-phase-averaged heralding probability, from Eq.(26),

$$P_{herald\,ave} = \frac{1}{4\pi} \int_{-\pi}^{\pi} d\phi \int_{0}^{\pi} \sin\chi \, d\chi \, P_{herald}. \tag{32}$$

**Figure 6** presents parametric plots of amplitude-and-phase-averaged memory-loading fidelity versus amplitude-and-phase-averaged heralding probability (averaging over 100 points on the Bloch sphere), while varying the cooperativity $C$, for four values of cavity loss rate $\kappa_J$. Fixed parameters values are $\Delta = 10\kappa$, $\sigma = \kappa/10$, and $\gamma = \kappa/10$. In all figures that follow we set the delay as $T = 1.2/\kappa$ and the interferometer phase as $\theta = \pi/2$ for *push-pull* and $\theta = 0$ for *on-off*. These values are very close to optimal in all cases considered and any further optimization affects the fidelities only at the 0.1% level. For the *push-pull* cases, the optimum value of $C$, that is $C_\pi$, is indicated by a large point (evaluated using Eq.(20), which is the point at which the phase-shift difference is exactly $\pi$ and which approximately maximizes the fidelity. The slight variation of $C_\pi$ from the maximum point results from the influence of the reflectivities ($R_1$, $R_2$) being less than 1. In all cases shown, the *push-pull* scheme outperforms the *on-off* scheme when asking for both the fidelity and heralding probability to be as large as possible. The trend shows that the presence of cavity loss reduces the achievable heralding probability at which maximum fidelity is achieved.

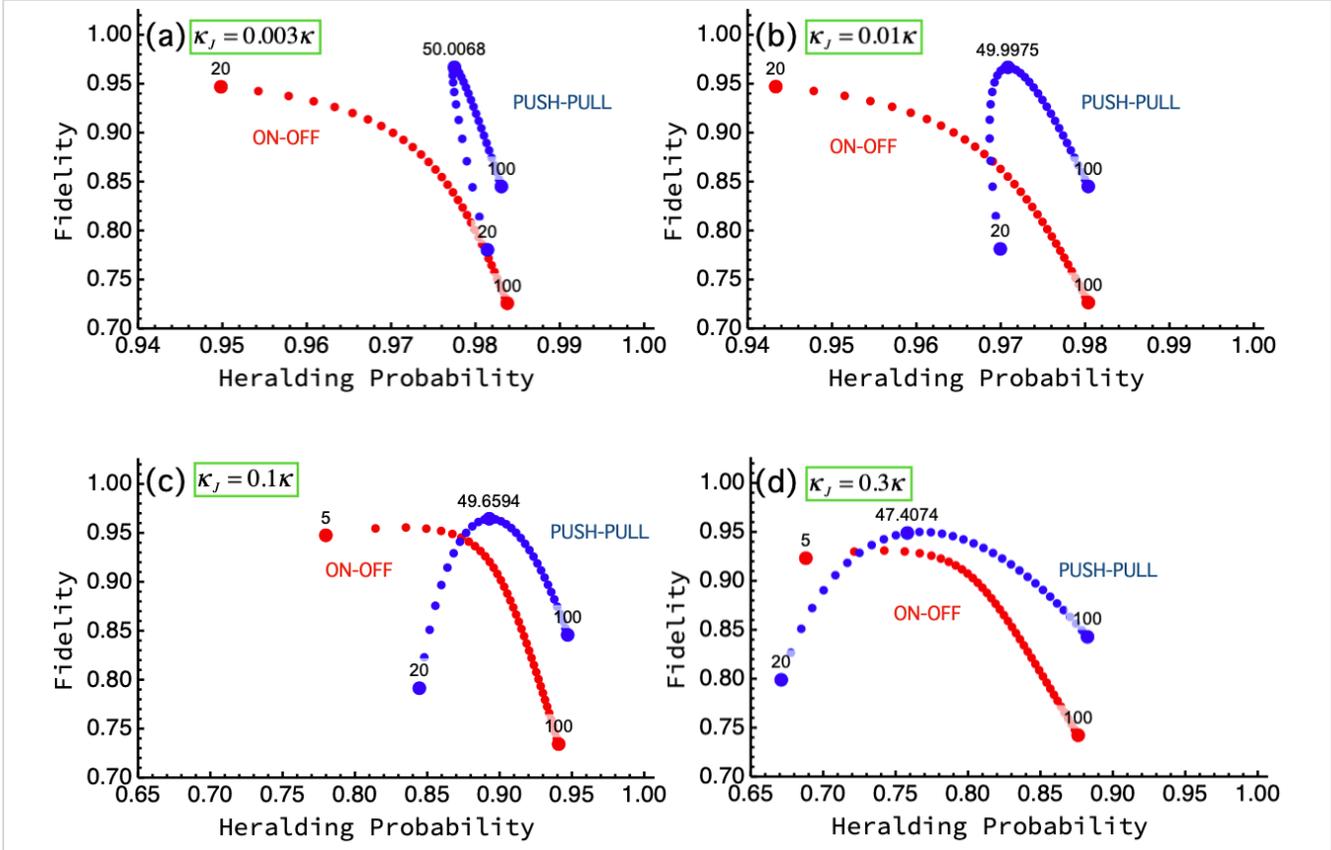

Fig. 6 Parametric plots of average memory-loading fidelity versus heralding probability, both averaged over the incoming single-photon polarization states, for four values of cavity loss rate $\kappa_J$. The *push-pull* results are shown in blue and *on-off* in red. Fixed parameter values are $\Delta = 10\,\kappa$, $\sigma = \kappa/10$, $\gamma = \kappa/10$. The cooperativity $C$ is varied along each trajectory of points, with extremal values of $C$ labeled at the terminating bold points. For the *push-pull* cases, the optimum value of $C$, that is $C_\pi$, is indicated by the number at a large intermediate bold point.

**Figure 7** shows a sequence of results with the cavity loss rate fixed at $\kappa_J = 0.03\kappa$ and the atomic level separation $\Delta$ varied. Fixed parameters are $\sigma = \kappa/10$, $\gamma = \kappa/10$, $\kappa_J = 0.003\kappa$. The results show that the *push-pull* scheme outperforms the *on-off* scheme even for decreasing level separation.

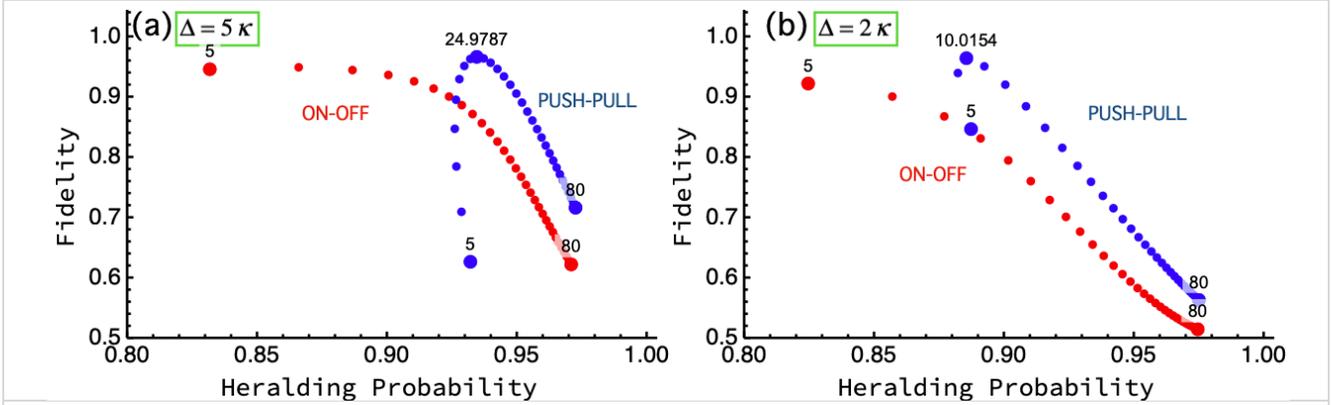

Fig. 7 Same as Fig. 6, with fixed cavity loss rate $\kappa_J$ and varied atomic level separation $\Delta$. For the *push-pull* plots, the values of $C_\pi$ are (a) 25, and (b) 10.0. C is varied from 5 to 80.

**Silicon vacancy (SiV) center**

An example of current interest is the silicon vacancy (SiV) center in a diamond photonic-crystal nanocavity, such as in [2,17,[22]], for which the four-state model is thought to be applicable for the spin-conserving transitions used. For a high-quality SiV nanocavity, parameters stated in these references can be approximated as $\Delta = 0.0043\,\kappa$, $\gamma = 0.00083\kappa$ and $g/\kappa = 0.050$ giving cooperativity $C \approx 13$. Assuming a relatively narrow-band input photon, $\sigma = \kappa/2000$, the present analysis yields the memory fidelity curves shown in **Fig. 8(a)**. We show in **Fig.8(b)** the hypothetical effect of decreasing the cavity loss by a factor of ten.

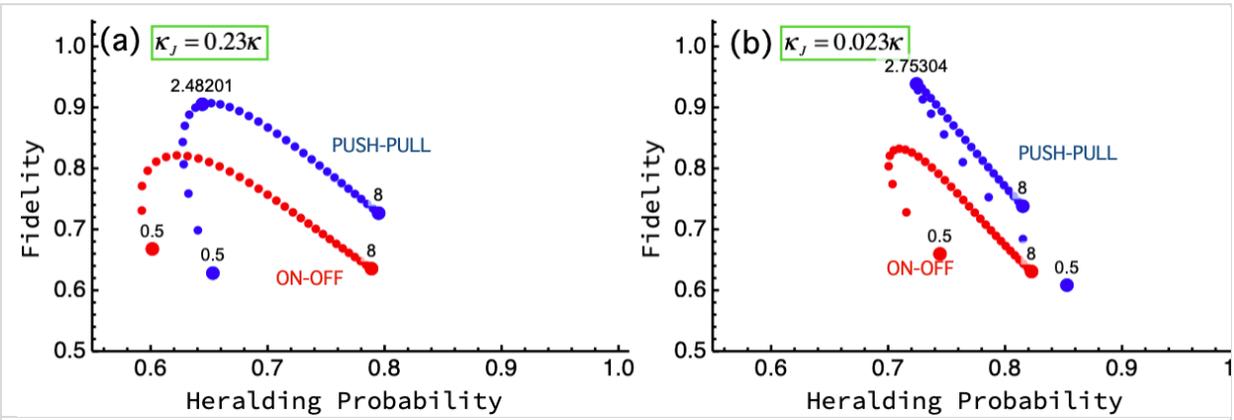

Fig. 8 (a) Same as Fig. 6, with parameters typical of those for an SiV nanocavity $\Delta = 0.0043\,\kappa$, $\gamma = 0.00083\kappa$, $\kappa_J = 0.23\kappa$ described in [22]. The photon linewidth is assumed to be $\sigma = \kappa/2000$. (b) Same as (a) but with cavity loss rate $\kappa_J$ ten times smaller, showing improved performance. For the *push-pull* plots, the values of $C_\pi$ are (a) 2.48, (b) 2.75. C varies from 0.5 to 8.0.

Also of interest is the effect of the incident photon bandwidth on the resulting memory fidelity. We show in **Fig. 9** a sequence of plots demonstrating that for this set of parameters the fidelity decreases with increasing photon bandwidth while the heralding probability increases.

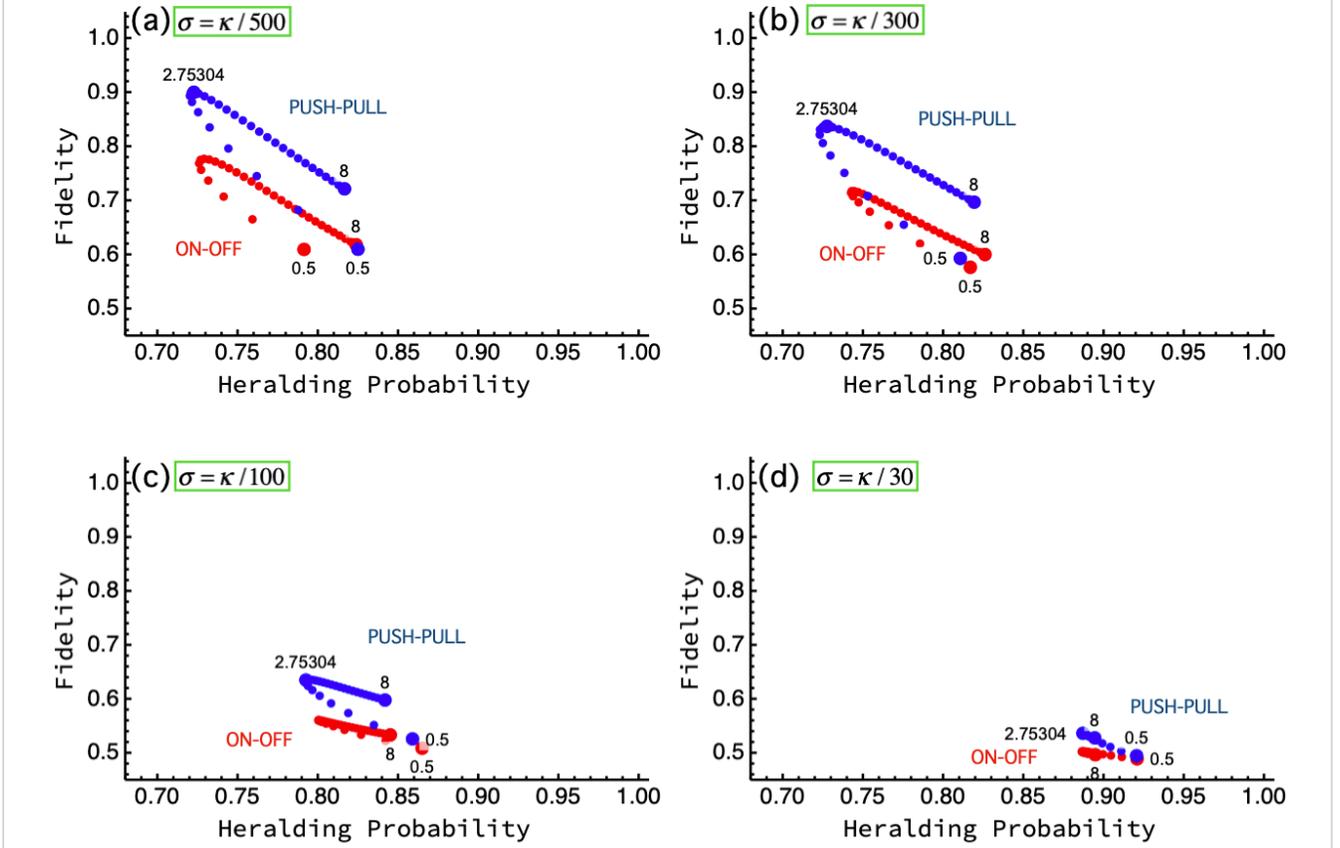

Fig. 9. Average memory fidelity vs. heralding probability for a hypthetical low-loss SiV nanocavity system. Parameters are the same as in Fig. 8(a). The bandwidth parameter of the incident photon is varied from $\kappa/500$ to $\kappa/30$.

**Discussion and Conclusions**

For loading a photonic qubit state into a cavity-atom system using variants of the Duan-Kimble scheme, the present analysis shows that for the considered parameter ranges the proposed *push-pull* scheme outperforms the original *on-off* scheme: it provides high memory-loading fidelity simultaneously with high heralding probability. The higher performance of the *push-pull* scheme likely occurs because at zero detuning it can provide equal cavity reflectivity for the two atomic states while providing phase-shift differences of exactly $\pi$. The results should play a role in optimizing future quantum repeater schemes based on the Duan-Kimble protocol.

For this analysis, we introduced a theoretical method that converts Heisenberg-picture equations of motion into Schrödinger-picture equations, thereby avoiding a theoretical error that appears in several prior publications. The method can be used to design a wide variety of schemes beyond the basic example treated here. For example, the cavity field can be bidirectional, and higher numbers of photons (and/or atoms) can be included. Several atom-cavity systems can be coupled together, as in [10].

## Acknowledgments

We thank Brian Smith, Franco Wong and Prajit Dhara for many helpful discussions on this topic, and in particular Franco for suggesting the name *push-pull* for our proposed detuning scheme. This work was supported by the Engineering Research Centers Program of the National Science Foundation under Grant #1941583 to the NSF-ERC Center for Quantum Networks.

## APPENDICES

### Appendix A Input-output theory

Here we present our derivation of the Schrödinger-picture equations of motion in the input-output formalism using standard methods. [9, 15, [23]] We include models for spontaneous emission damping as well as cavity loss (but not pure dephasing interactions with an environment). The unidirectional propagating photon-field operator (positive-frequency part) for the exterior field (that is, outside the cavity) is in the Heisenberg picture,

$$\hat{\Phi}^{(+)}(z,t) = \int_0^\infty \frac{dk}{2\pi} \hat{b}_k(t) e^{-i[\omega_0 + \omega(k)]t} e^{ikz}, \qquad (33)$$

where $\omega(k) = k/c - \omega_0$ is the frequency relative to the carrier $\omega_0$ and $k$ is the continuous variable for the propagation constant. Here $z < 0$ corresponds to the $\hat{A}(t)$ input field in Fig.4 and $z > 0$ corresponds to the $\hat{B}(t)$ output field. The mode operators satisfy $[\hat{b}_k(t), \hat{b}_{k'}^\dagger(t)] = 2\pi\delta(k-k')$.

For clarity we initially omit the $q$ cavity and the $J$ field channels and write the pure state for the atom, the cavity field $c$, and the combined exterior field ($A$ and $B$) as

$$|\psi(t)\rangle = \psi_e(t)|e\rangle_a|0\rangle_c|vac\rangle_{AB} + \psi_c(t)|g\rangle_a|1\rangle_c|vac\rangle_{AB} + \int_0^\infty \frac{dk}{2\pi} \psi_k(t)|g\rangle_a|0\rangle_c|1_k\rangle_{AB}, \qquad (34)$$

where the basis states are $|j\rangle_a|n\rangle_c|vac\rangle_{AB}$ or $|j\rangle_a|n\rangle_c|1_k\rangle_{AB}$, with $j = e,g$ for the atomic state, $n = 0,1$ for the cavity photon number, $|vac\rangle_{AB}$ indicating no photons in the exterior field and $|1_k\rangle_{AB}$ indicating one photon in the exterior field with wave number $k$. The Hamiltonian in the rotating-wave approximation is

$$\hat{H} = \hbar\tilde{\omega}_c \hat{c}^\dagger \hat{c} + \int_0^\infty \frac{dk}{2\pi} \hbar\omega_k \hat{b}_k^\dagger \hat{b}_k + \int_0^\infty \frac{dk}{2\pi} \hbar\left(if_k \hat{b}_k \hat{c}^\dagger - if_k \hat{b}_k^\dagger \hat{c}\right)$$
$$+ \hbar\omega_g |g\rangle\langle g| + \hbar\omega_e |e\rangle\langle e| - i\hbar g |e\rangle\langle g|\hat{c} + i\hbar g |g\rangle\langle e|\hat{c}^\dagger. \qquad (35)$$

where the exterior mode operators satisfy $[\hat{b}_k, \hat{b}_{k'}^\dagger] = 2\pi \delta(k-k')$, $\tilde{\omega}_c$ is the frequency of the 'bare' cavity mode (i.e. of a lossless cavity), not yet accounting for frequency pulling by the cavity damping mechanisms (see below). We take the coupling $f_k$ of the cavity mode to the exterior modes and the atom-field coupling $g$ as real (for the phase convention used here). The resulting Schrödinger equations of motion are

$$\partial_t \psi_e(t) = -i\omega_e \psi_e(t) + g\psi_c(t)$$
$$\partial_t \psi_c(t) = -i\tilde{\omega}_c \psi_c(t) + g\psi_e(t) + \int_0^\infty \frac{dk}{2\pi} f_k \psi_k(t) \qquad (36)$$
$$\partial_t \psi_k(t) = -i\omega_k \psi_k(t) - f_k \psi_c(t).$$

A standard derivation akin to the Wigner-Weisskopf approximation and summarized in **Appendix D**, eliminates the continuum of exterior modes to yield a damping term $\kappa$ and a frequency shift $\delta_c$ [19],

$$\partial_t \psi_c(t) = -(\kappa + i\omega_c)\psi_c(t) + g\psi_e(t) + F(t), \qquad (37)$$

where $\omega_c = \tilde{\omega}_c + \delta_c$, with

$$\kappa = \frac{1}{2c} f_{k(\tilde{\omega}_c)}^2, \qquad \delta_c = \frac{1}{c}\int_{-\omega_c}^\infty \frac{d\omega}{2\pi} f_{k(\omega)}^2 \frac{P}{\tilde{\omega}_c - \omega}, \qquad (38)$$

and $P$ indicating the principal value of the integral. The damping-induced frequency shift is an example of 'mode pulling.' The calculation relies on the assumption that the spectrum of the cavity field quasi-mode is narrow and strongly peaked at $\tilde{\omega}_c$ while the coupling to external modes is nearly constant and equals $f_{k(\tilde{\omega}_c)}$ in that spectral region. Using the fact that the fields are near-resonant with the cavity, the driving term is approximated as

$$F(t) \approx f_{k(\tilde{\omega}_c)} \int_0^\infty \frac{dk}{2\pi} \psi_k(t_p) e^{-i\omega_k(t-t_p)}, \qquad (39)$$

where $t_p$ is an arbitrarily chosen time in the distant past before the input interacts with the cavity. As we will see, $F(t)$ is proportional to the input field amplitude.

The standard way to identify the output field is to define it as the field at a time much later than the time at which the input pulse completes its interaction with the cavity, effectively creating a scattering theory. To find the form of the output field and its relation to the input, first solve the third line of Eq.(36) from a time $t_p$ in the distant past:

$$\psi_\omega(t) = \psi_\omega(t_p)e^{-i\omega(t-t_p)} - f_\omega \int_{t_p}^{t} dt' e^{-i\omega(t-t')} \psi_c(t') \ . \tag{40}$$

It is useful to form the quantity

$$\begin{aligned}\Xi(t) &= \int_0^\infty \frac{dk}{2\pi} f_k \psi_k(t) \\ &= \int_0^\infty \frac{dk}{2\pi} f_k \psi_k(t_p) e^{-i\omega_k(t-t_p)} - \int_{t_p}^{t} dt' \psi_c(t') \frac{1}{c} \int_{-\omega_c}^{\infty} \frac{d\omega_k}{2\pi} f_k^2 e^{-i\omega_k(t-t')}. \end{aligned} \tag{41}$$

The first term is approximated by replacing $f_k \to f_{k(\tilde{\omega}_c)}$ and defining the square-normalized input field amplitude as

$$E_{in}(t) = \int_0^\infty \frac{dk}{2\pi} \psi_k(t_p) e^{-i\omega_k(t-t_p)} \quad (t > t_p) \ , \tag{42}$$

The second term is evaluated using

$$\frac{1}{c} f_{k(\tilde{\omega}_c)}^2 \int_{-\omega_c}^{\infty} \frac{d\omega_k}{2\pi} e^{-i\omega_k(t-t')} \approx 2\kappa\, \delta(t-t') \ , \tag{43}$$

recognizing that this approximate delta function is to be used only against functions whose spectrum is narrow and strongly peaked at $\omega_0$. This gives

$$\Xi(t) \approx f_{k(\tilde{\omega}_c)} E_{in}(t) - \kappa \psi_c(t), \quad t > t_p \ . \tag{44}$$

Likewise, one can define a time $t_f$ in the far future and 'back-solve' the third line of Eq.(36) from $t_f$ to $t$,

$$\psi_\omega(t) = \psi_\omega(t_f)e^{-i\omega(t-t_f)} - f_\omega \int_{t_f}^{t} dt' e^{-i\omega(t-t')} \psi_c(t') \ . \tag{45}$$

Following the same steps as for $t_p$, one finds

$$\Xi(t) = f_{k(\tilde{\omega}_c)} E_{out}(t) + \kappa \psi_c(t), \quad t < t_f \ , \tag{46}$$

where the output field amplitude is defined as

$$E_{out}(t) = \int_0^\infty \frac{dk}{2\pi} \psi_k(t_f) e^{-i\omega_k(t-t_f)} . \tag{47}$$

Subtracting Eqs.(44) and (46) gives

$$-E_{out}(t) = -E_{in}(t) + (2\kappa / f_{k(\tilde{\omega}_c)}) \psi_c(t) . \tag{48}$$

Define the square-normalized input field as

$$A(t) = \frac{E_{in}(t)}{\sqrt{c}} = \int_{-\omega_c}^\infty \frac{d\omega}{2\pi} \frac{\psi_{k(\omega)}}{\sqrt{c}}(t_p) e^{-i\omega(t-t_p)} \doteq \int_{-\omega_c}^\infty \frac{d\omega}{2\pi} \tilde{A}(\omega) e^{-i\omega t} , \tag{49}$$

where $\tilde{A}(\omega) = c^{-1/2} \psi_{k(\omega)}(t_p) e^{i\omega t_p}$ and the (unnormalized) output field as

$$B(t) = -\frac{E_{out}(t)}{\sqrt{c}} = -\int_0^\infty \frac{d\omega}{2\pi} \frac{\psi_{k(\omega)}}{\sqrt{c}}(t_f) e^{-i\omega(t-t_f)} \doteq \int_{-\omega_c}^\infty \frac{d\omega}{2\pi} \tilde{B}(\omega) e^{-i\omega t} , \tag{50}$$

where $\tilde{B}(\omega) = -c^{-1/2} \psi_{k(\omega)}(t_f) e^{i\omega t_f}$. Using $f_{k(\tilde{\omega}_c)} = \sqrt{2\kappa}\sqrt{c}$, we arrive at

$$B(t) = -A(t) + \sqrt{2\kappa}\, \psi_c(t) , \tag{51}$$

as desired. Fourier transforming, we have

$$\tilde{B}(\omega) = -\tilde{A}(\omega) + \sqrt{2\kappa}\, \tilde{\psi}_c(\omega) , \tag{52}$$

as stated in Eq.(10). Revisiting Eq.(39) shows that

$$\begin{aligned}
F(t) &= \sqrt{2\kappa}\sqrt{c} \int_0^\infty \frac{d\omega/c}{2\pi} \psi_{k(\omega)}(t_p) e^{-i\omega(t-t_p)} \\
&= \sqrt{2\kappa}\sqrt{c} \int_0^\infty \frac{d\omega/c}{2\pi} \psi_k(t_0) e^{-i\omega(t-t_p)} \\
&= \sqrt{2\kappa} \int_0^\infty \frac{d\omega}{2\pi} \tilde{A}(\omega) e^{-i\omega t} = \sqrt{2\kappa}\, A(t).
\end{aligned} \tag{53}$$

Thus, the cavity equation is

$$\partial_t \psi_c(t) = -(\kappa + i\omega_c)\psi_c(t) + g\psi_e(t) + \sqrt{2\kappa}\, A(t) . \tag{54}$$

Finally, including the $Q$ cavity and the $J$ field channels and repeating the above steps leads to Eq.(7).

**Appendix B Optimizing the *on-off* phase shifts**

To derive Eqs. (22) and (23), start with Eq.(17),

$$r_j(0) = \frac{(1+i\Delta_j/\gamma)(1-\kappa_J/\kappa)-C}{(1+i\Delta_j/\gamma)(1+\kappa_J/\kappa)+C}, \tag{55}$$

giving the amplitude reflection coefficient when the photon is very narrow-band and tuned to the cavity resonance. For the *on* case, we take the atom-cavity detuning $\Delta_1 = 0$, and thus

$$r_{ON} = \frac{(1-\kappa_J/\kappa)-C}{(1+\kappa_J/\kappa)+C}. \tag{56}$$

Then for $\kappa_J/\kappa \ll 1$ we have $r_{ON} = -1$. For the *off* case, we take $\Delta_2 \neq 0$, and assume it to be large ($\Delta_2 \gg \gamma$) to minimize the atom's effect on the field. To make the phase difference between *on* and *off* cases as close as possible to $\pi$, one needs to minimize the magnitude of the quantity

$$|r_{OFF}(0)-1| = \left|2\frac{(1+i\Delta_2/\gamma)(\kappa_J/\kappa)+C}{(1+i\Delta_2/\gamma)(1+\kappa_J/\kappa)+C}\right|$$
$$= \sqrt{\frac{(\Delta_2/\gamma)^2(\kappa_J/\kappa)^2+C^2}{(\Delta_2/\gamma)^2(1+\kappa_J/\kappa)^2+C^2}}. \tag{57}$$

To make this quantity much smaller than 1 requires $(\Delta_2/\gamma)(1+\kappa_J/\kappa) \gg C$. Then if $(\Delta_2/\gamma)(\kappa_J/\kappa) \ll C$, we need $(\Delta_2/\gamma)^2 \gg C^2$.

Under these conditions, the phase-shift error $\delta_{error}$ (departure of $\delta_{phase}(0)$ from $\pi$) for the *off* condition is estimated by considering the real and imaginary parts, $r_{OFF}(0) = X + iY$, with phase-difference error estimated in a small-approximation as $\delta_{error} \approx Y/X$, evaluated to be

$$\delta_{error} = \frac{Y}{X} \approx \frac{2C(\Delta_2/\gamma)}{(\Delta_2/\gamma)^2(1-\kappa_J^2/\kappa^2)-2C\kappa_J/\kappa-C^2} \approx \frac{2C(\Delta_2/\gamma)}{(\Delta_2/\gamma)^2(1-\kappa_J^2/\kappa^2)}, \tag{58}$$

where we used $2C\kappa_J/\kappa + C^2 \ll (\Delta_2/\gamma)^2$.

**Appendix C Loading the photon state into the memory**

The normalized incoming photon state before the PBS is $|\psi(0)\rangle = \alpha|\varphi\rangle_H + \beta|\varphi\rangle_V$, where $\varphi$ indicates the temporal mode of the photon, which is the same for the two polarization states $H$ and $V$ and defined by $\tilde{A}(\omega)$ as

$$|\varphi\rangle_H = \int \frac{d\omega}{2\pi} \tilde{A}(\omega)\hat{a}_H^\dagger(\omega)|vac\rangle$$
$$|\varphi\rangle_V = \int \frac{d\omega}{2\pi} \tilde{A}(\omega)\hat{a}_V^\dagger(\omega)|vac\rangle, \quad (59)$$

with

$$\int \frac{d\omega}{2\pi} |\tilde{A}(\omega)|^2 = 1$$
$$[\hat{a}_p(\omega), \hat{a}_q^\dagger(\omega')] = 2\pi \delta(\omega-\omega')\delta_{pq}. \quad (60)$$

The overall state after the PBS, but before cavity reflection, is

$$|\psi\rangle = \left(\alpha|\varphi\rangle_{UH} + \beta|\varphi\rangle_{LV}\right) 2^{-1/2}\left(|g_1\rangle + |g_2\rangle\right)_{atom}, \quad (61)$$

where subscripts $U$ and $L$ label upper and lower paths in **Fig. 2**. The action of the cavity reflection is, depending on the state of the atom (1 or 2),

$$|\varphi\rangle_{UH} \to |\varphi_{1,2}\rangle_{UH} = \int \frac{d\omega}{2\pi} \varphi_{1,2}(\omega)\hat{a}_{UH}^\dagger(\omega)|vac\rangle, \quad (62)$$

where the (un-normalized) state amplitude is

$$\varphi_j(\omega) \doteq r_j(\omega)\tilde{A}(\omega) \quad (j=1,2), \quad (63)$$

with $r_{1,2}(\omega)$ being the complex reflectivity. The state in the lower path undergoes adjustable phase and time delays represented by $e^{i\theta}e^{i\omega T}$,

$$|\varphi\rangle_{LV} \to |\varphi_{\theta T}\rangle_{LV} = \int \frac{d\omega}{2\pi} \varphi_{\theta T}(\omega)\hat{a}_{LV}^\dagger(\omega)|vac\rangle, \quad (64)$$

where

$$\varphi_{\theta T}(\omega) \doteq e^{i\theta}e^{i\omega T}\tilde{A}(\omega). \quad (65)$$

After polarization rotation in the upper path, the state (un-normalized) is

$$2^{-1/2}\alpha\left(|\varphi_1\rangle_{UV}|g_1\rangle+|\varphi_2\rangle_{UV}|g_2\rangle\right)+\beta|\varphi_{\theta T}\rangle_{LV}\left(|g_1\rangle+|g_2\rangle\right)$$

$$2^{-1/2}\int\frac{d\omega}{2\pi}\left\{\alpha\left(\varphi_1(\omega)|g_1\rangle+\varphi_2(\omega)|g_2\rangle\right)\hat{a}_{UV}^\dagger(\omega)+\beta\varphi_{\theta T}(\omega)\left(|g_1\rangle+|g_2\rangle\right)\hat{a}_{LV}^\dagger(\omega)\right\}|vac\rangle. \tag{66}$$

To balance the interferometer, one could include an optional attenuation factor $\eta$ in the lower path by replacing $\beta \to \eta^{1/2}\beta$, although we have not studied this modification in detail.

The beam splitter combines the paths into paths leading to detectors labeled $D_s$ (where $s=\pm 1$ labels which detector path the photon takes), according to

$$\hat{a}_{UV}^\dagger(\omega)=2^{-1/2}\left(\hat{a}_{+1}^\dagger(\omega)+\hat{a}_{-1}^\dagger(\omega)\right)$$
$$\hat{a}_{LV}^\dagger(\omega)=2^{-1/2}\left(\hat{a}_{+1}^\dagger(\omega)-\hat{a}_{-1}^\dagger(\omega)\right). \tag{67}$$

The system is then described by the state

$$2^{-1}\sum_{s=\pm 1}\int\frac{d\omega}{2\pi}\left\{\left(\alpha r_1(\omega)+s\beta e^{i\theta}e^{i\omega T}\right)|g_1\rangle+\left(\alpha r_2(\omega)+s\beta e^{i\theta}e^{i\omega T}\right)|g_2\rangle\right\}\tilde{A}(\omega)\hat{a}_s^\dagger(\omega)|vac\rangle. \tag{68}$$

An inverse Hadamard transformation is applied to the atomic qubit, leaving the (unnormalized) state prior to detection to be

$$|\psi'\rangle_s=2^{-1}2^{-1/2}\int\frac{d\omega}{2\pi}\left\{\phi_1(\omega)|g_1\rangle+\phi_{2s}(\omega)|g_2\rangle\right\}\tilde{A}(\omega)\hat{a}_s^\dagger(\omega)|vac\rangle, \tag{69}$$

where

$$\phi_1(\omega)=\alpha\left(r_1(\omega)-r_2(\omega)\right),\quad \phi_{2s}(\omega)=\left(\alpha\left[r_1(\omega)+r_2(\omega)\right]+2s\beta e^{i\theta}e^{i\omega T}\right). \tag{70}$$

In practice, the inverse Hadamard is applied after a detection event, but this makes no difference to the result.

As a simple example, consider the ideal case in which a $\pi$ phase difference occurs between the reflectivities from the two atomic states and that the temporal mode is not distorted on reflection, $\varphi_1(\omega)=-\varphi_2(\omega)=\varphi(\omega)$. Then, up to an overall phase,

$$|\psi'\rangle_s=2^{-1/2}\left\{\alpha|g_1\rangle+s\beta|g_2\rangle\right\}|\varphi\rangle_s, \tag{71}$$

where the factor $2^{-1/2}$ indicates the probability amplitude for occurrence of a particular outcome labeled by $s$. Depending on which detector clicks ($s=\pm 1$), a unitary operation on the atomic qubit ($|g_2\rangle \to -|g_2\rangle$) may or may not be needed to leave the memory in the targeted state $\alpha|g_1\rangle + \beta|g_2\rangle$.

Given a detection event at the $s$ detector, the state of the atom is conditioned (updated) to a unnormalized mixed state represented by the density matrix

$$2^{-2}2^{-1}\int \frac{d\omega}{2\pi} |\tilde{A}(\omega)|^2 \{\phi_1(\omega)|g_1\rangle + \phi_{2s}(\omega)|g_2\rangle\}\{\phi_1^*(\omega)\langle g_1| + \phi_{2s}^*(\omega)\langle g_2|\}. \tag{72}$$

Normalizing the state ($\text{Tr}\,\hat{\rho}_s = 1$) gives

$$\hat{\rho}_s = \frac{1}{8K_s}\int \frac{d\omega}{2\pi} |\tilde{A}(\omega)|^2 \{\phi_1(\omega)|g_1\rangle + \phi_{2s}(\omega)|g_2\rangle\}\{\phi_1^*(\omega)\langle g_1| + \phi_{2s}^*(\omega)\langle g_2|\}, \tag{73}$$

where

$$K_s = \frac{1}{8}\int \frac{d\omega}{2\pi} |\tilde{A}(\omega)|^2 \left\{|\alpha(r_1(\omega) - r_2(\omega))|^2 + |\alpha[r_1(\omega) + r_2(\omega)] + 2s\beta e^{i\theta}e^{i\omega T}|^2\right\}. \tag{74}$$

The fidelity is calculated to be

$$\begin{aligned}
F_s &= \{\alpha^*\langle g_1| + s\beta^*\langle g_2|\}\hat{\rho}_s\{\alpha|g_1\rangle + s\beta|g_2\rangle\} \\
&= \frac{1}{8K_s}\int \frac{d\omega}{2\pi} |\tilde{A}(\omega)|^2 |\alpha^*\phi_1(\omega) + s\beta^*\phi_{2s}(\omega)|^2 \\
&= \frac{1}{8K_s}\int \frac{d\omega}{2\pi} |\tilde{A}(\omega)|^2 \left|\left(|\alpha|^2 + s\beta^*\alpha\right)r_1(\omega) - \left(|\alpha|^2 - s\beta^*\alpha\right)r_2(\omega) + 2|\beta|^2 e^{i\theta}e^{i\omega T}\right|^2.
\end{aligned} \tag{75}$$

In the ideal case, that is $r_1(\omega) = -r_2(\omega) = e^{i\theta}e^{i\omega T}$ over the spectral range of $\tilde{A}(\omega)$, the results reduce to $K_s = 1/2$ and $F_s = 1$.

(76)

The total probability for a detection event (at either detector) is given by

$$\begin{aligned}
P_{detect} &= K_{+1} + K_{-1} \\
&= |\beta|^2 + |\alpha|^2 \left(\frac{R_1 + R_2}{2}\right),
\end{aligned} \tag{77}$$

where the integrated reflection probabilities are

$$R_j = \int \frac{d\omega}{2\pi} |\tilde{A}(\omega)|^2 |r_j(\omega)|^2. \tag{78}$$

# Appendix D Elimination of the continuum of exterior modes

In **Appendix A** we invoked a standard method to eliminate the continuum of exterior modes to yield a damping term. [19] In brief, solve the third line in Eq.(36) from $t_0$ and insert into the second line for $\partial_t \psi_c(t)$, defining a slowly varying variable by

$$\bar{\psi}_c(t') = e^{i\tilde{\omega}_c t'} \psi_c(t') . \tag{79}$$

Change integration variable to $\tau = t - t'$, define $\bar{t} = t - t_p$, and use $k = (\omega_c + \omega_k)/c$ to evaluate one of the resulting terms as

$$\int_0^\infty \frac{dk}{2\pi} f_k^2 \int_0^{\bar{t}} d\tau e^{-i\omega_k \tau} \bar{\psi}_c(t-\tau) e^{-i\tilde{\omega}_c(t-\tau)}$$

$$\approx e^{-i\tilde{\omega}_c t} \bar{\psi}_c(t) \int_0^{\bar{t}} d\tau \, G(\tau) \tag{80}$$

where $G(\tau) = \int_{-\omega_c}^\infty \frac{d\omega/c}{2\pi} f_{k(\omega)}^2 e^{-i(\omega - \tilde{\omega}_c)\tau}$ .

on the assumption that $\bar{t}$ is sufficiently large to make $G(\tau)$ decays to zero rapidly as $\tau$ increases from zero and that $\bar{\psi}_c(t-\tau)$ is slowly varying compared to rate at which $G(\tau)$ decays. The $\omega$ integral is evaluated to give the standard decay rate and cavity frequency shift in Eq.(38).

The other term in $\partial_t \psi_c(t)$ equals

$$F(t) = \int_0^\infty \frac{dk}{2\pi} f_k \psi_k(t_p) e^{-i\omega_k(t-t_p)} , \tag{81}$$

which is approximated as in Eq.(39).